# Self-powered Filterless On-chip Full-Stokes Polarimeter

Chen Fang[1], Junze Li[1], Boxuan Zhou[1], and Dehui Li[1, 2,]*

[1]School of Optical and Electronic Information, Huazhong University of Science and Technology, Wuhan, 430074, China

[2]Wuhan National Laboratory for Optoelectronics, Huazhong University of Science and Technology, Wuhan, 430074, China

*Corresponding author. Email: dehuili@hust.edu.cn.

**Abstract:**

The detection of polarization states of light is essential in photonic and optoelectronic devices. Currently, the polarimeters are usually constructed with the help of waveplates or a comprehensive metasurface, which will inevitably increase the fabrication complexity and unnecessary energy loss. Here, we have successfully demonstrated a self-powered filterless on-chip full-Stokes polarimeter based on a single-layer $MoS_2$/few-layer $MoS_2$ homojunction. Combining the built-in electric field enhanced circular photogalvanic effect with the intrinsic optical anisotropy of $MoS_2$ between in-plane and out-of-plane direction, the device is able to conveniently sense four Stokes parameters of incident light at zero bias without requiring an extra filtering layer, and can function in the wavelength range of 650-690 nm with acceptable average errors. Besides, this homojunction device is easy to integrate with silicon-based chips and could have much smaller sizes than metasurface based polarimeters. Our study thus provides an excellent paradigm for high-performance on-chip filterless polarimeters.

*dichalcogenides, homojunction*

**Introduction**

Polarization reflects the vectorial characteristic of light, which plays an increasingly important role in various applications including astronomy, remote sensing, military detection, medical diagnosis and optical communication.[1-8] Although studies on polarization devices have so far made considerable achievements, probing states of polarization (SOP) is still a challenge in integrated on-chip polarization photonic systems.[9,10] Traditional polarization analyzers or full-Stokes polarimeters often contain complex and bulky optical systems including various polarizers and waveplates, which greatly hinders to miniaturize the photonic devices and develop the integrated photonic chips. In order to solve this problem, a number of on-chip polarization detectors which are consisted by an ultra-compact structure and high integration density have been developed in recent years.[1,3,4,9-13] Such full-Stokes polarimeters either rely on coupling a polarization-unresolved photodetector with a comprehensive plasmonic metasurface/grating or using a separated micro-retarder layer on top of a linear polarization layer for the extraction of different polarization components.[9,11-13] Although those on-chip polarimeters can easily integrate with photonic chips and enable more functionalities such as measurements of wavelength, intensity as well as spatial distribution,[4,9,14] the additional filtering layers would not only increase the complexity of the system and device cost, but also cause unnecessary energy loss, which is unfavorable in photonic devices and integrated photonic chips. To this end, it is necessary to further simplify the device structure and explore new filterless full-Stokes polarimeters for integrated optoelectronic and photonic applications.

Two-dimensional (2D) layered materials have been extensively investigated for their atomically thin thickness, layered structure and unique optical and optoelectronic properties.[15-18] Particularly, the layered nature of 2D materials allows them to form various van der Waals junctions with functionalities according to demanding, and easily integrate with silicon-based chips without requiring the lattice mismatch.[15,18,19] 2D materials with in-plane anisotropy such as black phosphorus and $ReS_2$ have been utilized to detect the linearly polarized light with decent performance.[2,20,21] Nevertheless, up to date, sensing all polarization states of light based on 2D materials remains elusive. The main obstacle lies in the detection of circularly polarized light components partially due to the lack of 2D materials that can efficiently distinguish the right-handed (σ+) or left-handed (σ−) circularly polarized light.

Single-layer (SL) transition metal dichalcogenides (TMDCs) show promising potential for circularly polarized light detection taking advantage of their special band structure where two nonequivalent valleys appear in the K-space of the Brillouin zone and thus leads to the valley-dependent optical selection rules.[22-24] The valley-dependent optical selection rule can induce the valley-locked spin-polarized photocurrent which is called circular photogalvanic effect (CPGE).[25-28] With the aid of CPGE in SL TMDCs, the helicity of the circularly polarized light can be identified. Nevertheless, this effect is usually rather weak and only high-intensity excitation can generate a detectable CPGE current,[25,26,28] which makes it difficult to detect weak circularly polarized light. Applying a bias to the device can significantly improve the CPGE response;[27,28] nevertheless, this will also increase the noise current and thus deteriorate the device performance. Therefore, it is urgent to explore new approaches for improving the CPGE current under low incident power density.

2D van der Waals junctions is another possible solution, which could enhance the CPGE current via the built-in electric field in the junctions.[29-31] Recent study reveals that obvious circular photocurrent has been observed in $MoS_2/WSe_2$ heterojunction under a normal incidence, confirming the modulation effect of van der Waals junctions on the CPGE current.[31] In this sense, it is feasible to detect all SOP of light in a SL TMDC based van der Waals junction by utilizing its CPGE effect and intrinsic optical anisotropy between in-plane and out-of-plane direction. Here, we have successfully demonstrated a self-powered filterless on-chip full-Stokes polarimeter in a SL-$MoS_2$/few-layer $MoS_2$ (FL-$MoS_2$) homojunction by using the built-in electric field enhanced CPGE current and the linearly polarized photoresponse between in-plane and out-of-plane direction of $MoS_2$.

**Results and discussion**

*Enhanced CPGE response through built-in electric field*

Figure 1a presents the schematic illustration of the SL-$MoS_2$/FL-$MoS_2$ homojunction device. Both SL-$MoS_2$ and FL-$MoS_2$ flakes were obtained through mechanical exfoliation from $MoS_2$ bulk crystals and stacked to form the homojunction by a dry transfer method. The energy band diagram of the homojunction reveals that the holes in the valence band of SL-$MoS_2$ can easily transfer to FL-$MoS_2$ while the electron transfer can be neglected due to the similar electron affinity between SL-$MoS_2$ and FL-$MoS_2$.[32] The optical microscope (OM) and fluorescence images of the device are shown in Figure 1b. It is obvious that the constitute SL-$MoS_2$ exhibits bright red luminescence whereas its intensity sharply decreases in the junction region, which indicates the efficient charge transfer in the homojunction and thus the excellent interlayer coupling between SL-MoS$_2$ and FL-MoS$_2$ flakes. The Raman spectra of the device in Figure 1c show two distinct scattering peaks at around 380 and 400 cm$^{-1}$, which corresponds to E$^1_{2g}$ and A$_{1g}$ modes, respectively. The difference between the two peaks is 19.1 cm$^{-1}$ for SL-MoS$_2$ while increases to 24.1 cm$^{-1}$ for FL-MoS$_2$, confirming that the flakes we used are single layer and few layer, respectively.[33] It is clear that the peaks of SL-MoS$_2$ and FL-MoS$_2$ co-exist in the overlapped region, which proves the formation of homojunction and the SL-MoS$_2$ can retain its intrinsic properties after stacking the junction (Figure 1c). Figure 1d presents the PL spectra in different regions of the device. The obvious PL quenching effect in the junction region further demonstrates the efficient separation of photogenerated carriers induced by the built-in electric field.

We have further analyzed the CPGE and optical anisotropy in SL-MoS$_2$ at room temperature, which is essential for detecting the SOP of light by using SL-MoS$_2$ devices. As shown in Figure S1, on-resonance circularly polarized light excitations can result in a light-induced valley population imbalance and thus give rise to the carrier net momentum between the two valleys. Then a CPGE photoresponse is generated even without applying any bias.[27,28,34-37] Through a λ/4 plate modulating the incident light polarization, the total photocurrent as a function of the λ/4 plate angle $\varphi$ can be described phenomenologically by:

$$I = C\sin 2\varphi + L_1\sin 4\varphi + L_2\cos 4\varphi + A$$

where $C$ is the coefficient of the CPGE current and $L_1$, $L_2$ are related to the linear photocurrent such as linear photogalvanic effect (LPGE) current while $A$ is the polarization-independent offset current.[26,27,34] In general, the magnitude of the CPGE current is related to the second-order tensor which is determined by the structure symmetry of a material.[28] Since most TMDC components have an in-plane isotropic structure, they usually present a distinct CPGE response only under oblique incident excitations.[28] Particularly, SL TMDCs also exhibit a linearly polarized response under such oblique incident conditions due to the optical anisotropy between in-plane and out-of-plane direction of SL TMDCs.[22] Thus, it is possible to sense all SOP of light by using SL TMDCs under oblique incidence.

We have designed a convenient SOP measurement device base on the SL-MoS$_2$/FL-MoS$_2$ homojunction (Figure 2a). The SOP of the incident monochromatic light was modulated by a polarizer, a λ/2 plate coupled with a λ/4 plate and then the light was perpendicularly focused onto the homojunction device, which was aslant fixed on a rotation stage at a 45° angle to the horizontal plane. Thus the monochromatic light is equivalent to be oblique incident on the sample under an angle of 45°. The generated photocurrent was amplified by a pre-amplifier and measured by a lock-in amplifier.

Figure 2b shows the helicity-dependent photocurrent of the as-fabricated homojunction device at zero bias under an excitation power of 60 μW cm$^{-2}$. The photocurrent varies significantly with the λ/4 plate angle under different excitation wavelengths, suggesting that our homojunction can distinguish the helicity of the circularly polarized light within a wide wavelength range. According to the discussion above, CPGE current is generally weak and a high-intensity laser excitation is required to observe noticeable CPGE current, which is not conducive to the photodetection. Therefore, we employed a SL-MoS$_2$/FL-MoS$_2$ homojunction device, in which CPGE current can be significantly enhanced and a strong CPGE signal can be generated under a relatively weak excitation compared with that in a SL-MoS$_2$ device. To clearly see the enhancement of the CPGE current in the homojunction device, the helicity-dependent photocurrent of both the homojunction device and the SL-MoS$_2$ are displayed in Figure 2c. The photocurrent for the homojunction device was acquired at a zero bias while a small bias was applied for SL-MoS$_2$ device since the CPGE is too weak to be detected at the same excitation in SL-MoS$_2$ device under a zero bias. It is obvious that the homojunction device greatly improves the photocurrent and presents a larger response difference under σ−/σ+ circularly polarized light excitation even at zero bias. Although some other photogeneration mechanisms such as linear photon drag effect (LPDE) can also be enhanced by the built-in electric field, they do not contribute to the circular response, which confirms the enhancement of the CPGE current in SL-MoS$_2$. In addition, we also extracted the associated CPGE coefficient |C| in Figure 2d to further illustrate the enhanced CPGE response of the homojunction, and prove its high polarization resolution ability for homojunction based photodetectors.

The possible underlying mechanism of this enhanced CPGE current in homojunction device under on-resonance and quasi-resonance excitation is schematically shown in Figure 2e. Owing to the valley-dependent optical selection rule in SL-MoS$_2$, σ+ or σ− circularly polarized light can only excite the carriers at the K or K′ valley, which thus leads to the valley population imbalance.[24,26] Under illumination by a circularly polarized light with certain helicity, the resulting CPGE current is a superposition of electron current and hole current, which propagates along same direction but with an opposite sign.[31,38] Therefore, electron current and hole current tend to cancel each other in each single valley. In this SL-MoS$_2$/FL-MoS$_2$ homojunction, the special band alignment enables the holes to undergo ultrafast transfer from SL-MoS$_2$ to the FL-MoS$_2$ due to the relative large valence band offset between them, while the electron transfer can be neglected. Owing to the conservation of the inversion symmetry in FL-MoS$_2$, the transferred holes cannot generate CPGE current.[39] As a result, the hole current is significantly reduced and the net CPGE current in the homojunction mainly originates from electron current, which is thus greatly enhanced in such SL-MoS$_2$/FL-MoS$_2$ homojunction, similar to that in a MoS$_2$/WSe$_2$ heterostructure.[31] In addition, the built-in electric field in such homojunction would also expect to contribute to the enhanced CPGE current.[27,28]

*SL-MoS$_2$/FL-MoS$_2$ homojunction full-Stokes polarimeter*

In particular, under the oblique illumination, by using the optical anisotropy between the in-plane and out-of-plane direction, an intrinsic linear polarization response would also be introduced in SL-MoS$_2$/FL-MoS$_2$ homojunction. Assisted by the built-in electrical field, a linear dichroic ratio of 1.2 was achieved at zero bias under various excitation wavelength (Figure 2f). Equipped with both linear polarization and circular polarization discrimination capability under low incident power density, SL-MoS$_2$/FL-MoS$_2$ homojunction device would expect to be able to sense all SOP of light without requiring an external bias.

To accurately characterize the polarized light, four Stokes parameters ($S_0$, $S_1$, $S_2$, and $S_3$) are normally used to completely describe the SOP of light.[4,9] The linear Stokes vector $S_1$ and $S_2$ characterize the intensity of the horizontal/vertical or 45° linear polarization component, while the parameter $S_3$ represents the intensity of the σ+ or σ− circular polarization component. The parameter $S_0$ defines the total incident power. For our as-fabricated full-Stokes polarimeter, the measured parameters $S_0$, $S_1$ and $S_2$ mainly depend on the linear polarization-sensitive photocurrent while the parameter $S_3$ is largely determined by the CPGE response (Figure S2). Therefore, our homojunction device can be viewed as the integration of polarizer, λ/2 plate, λ/4 plate and a polarization independent photodetector together. In detail, under a certain incident wavelength and power density, the device calibration parameters $I_M$, $I_m$, $I_l$ and $I_r$ should be first measured, where $I_M$ and $I_m$ are the maximum and minimum photocurrent of the linear polarization response while $I_l$ and $I_r$ are the circular response under σ− or σ+ circularly polarized light excitation. Afterwards, in order to obtain the SOP of incident light, we need to acquire photocurrents $I_0$, $I_{45}$, $I_{90}$, $I_{135}$ and $I_{180}$ under rotation angles of 0°, 45°, 90°, 135° and 180° with the zero-degree of orientation setting as the polarization orientation of $I_M$. Then all-Stokes parameters of the incident light can be extracted based on the model we proposed (Supplementary text). It should be noted that when the excitation wavelength or power changes, the device parameters $I_M$, $I_m$, $I_l$ and $I_r$ need to be recalibrated to ensure the correct results.

According to this model, we further extract the device calibration parameters and verify the performance of our homojunction device. The calibration parameters $I_M$, $I_m$, $I_l$ and $I_r$ within the wavelength range from 650 to 690 nm with a power of 60 μW cm$^{-2}$ are first extracted from Figure 2b, f and displayed in Figure 3a. To be a full-Stokes polarimeter, the detector should accurately resolve the SOP of light with various unknown polarization states. We then acquire the photocurrent $I_0$, $I_{45}$, $I_{90}$, $I_{135}$ and $I_{180}$ of the incident light with a wavelength of 670 nm and eleven different SOPs including linear polarization, circular polarization, elliptic polarization and partial polarization generated by the polarization modulation module to verify our polarimeter. The acquired photocurrents for different incident light are presented in Figure 3b, c. Obviously the measured photocurrent varies significantly with the rotation angle, which is mainly due to the different linear polarization response of the homojunction. Base on the calibration parameters (Figure 3a) and the acquired photocurrents under different rotation angles (Figure 3b,c), we extracted the normalized Stokes parameters based on the computation model we proposed (Figure 3d). For the comparison, we also displayed the corresponding given Stokes parameters in Figure 3d, which are calculated based on the SOP of the incident light. It is obvious that all the components of measured Stokes parameters agree well with the given parameters for all input light we used.

In addition to the 670 nm light excitation, the device also exhibits polarization discrimination capability within the wavelength range from 650 nm to 690 nm. The measured rotation angle-dependent photocurrents under 650, 660, 680 and 690 nm irradiation are shown in Figure S3 for several input light excitations with different polarization. Following the same measurement and computation procedure as the 670 nm light excitation, we have extracted Stokes parameters under different wavelength light excitation, shown in Figure S4 together with the input values. Overall, our device can also measure Stokes parameters with a small margin of errors in the wavelength range. To evaluate the measurement deviation under different incident conditions, we have calculated the arithmetical mean errors of the three parameters $S_1$, $S_2$, $S_3$ at different polarized light excitation (Figure 3e).[9] It is worthy noticing that the evaluated results are more accurate in terms of parameters $S_1$ and $S_2$ while the errors of parameters $S_3$ is relatively large regardless of the excitation wavelength, which can be attributed to the limited ability of our homojunction to distinguish circularly polarized light. In addition, the measurement errors at 670 nm is smaller than that of other wavelengths, especially for parameter $S_3$, which is mainly due to the higher degree of circular photocurrent under on-resonance excitation.[26,28]

Remarkably, the minimum errors of the device are 5%, 4.8% and 6.7% for parameters $S_1$, $S_2$ and $S_3$, respectively, thus presenting good reliability among the current full-Stokes polarimeter (Table S1).[3,9,11,40,41]

We have also investigated the incident power density dependent performance of the homojunction device under the 670 nm light excitation (Figure S5 and Figure S6). The Stokes parameters can still be evaluated within the excitation power range we investigated with a reasonable error. Nevertheless, as the incident power decreases, the lower polarization response will increase the measurement error correspondingly, especially the circular polarization parameter $S_3$ (Figure 3f). In addition to the homojunction devices, the polarization states of light can also be measured by using a SL-MoS$_2$ device (Figure S7), which would greatly reduce the complexity of the device fabrication process. Nevertheless, under the same excitation with an incident power of 60 μW cm$^{-2}$, the device exhibits much lower photoresponse and CPGE current than that in SL-MoS$_2$/FL-MoS$_2$ homojunction, thus resulting in the considerable errors in full-Stokes parameter detection (Figure S7e). In addition, a low bias is usually required in order to achieve an observable liner polarization response in SL-MoS$_2$ device. Therefore, the homojunction device exhibits remarkably improved performance under a low-intensity excitation and can function perfectly well as a self-powered full-Stokes polarimeter.

We have finally evaluated the photodetection capability of the homojunction device via photoconductivity measurement under unpolarized excitation (Figure 4), which is essential for the polarization detectors since it can determine the detection efficiency and the signal extraction ability.[10] Figure 4a displays the *I-V* curves of the as-fabricated homojunction both in dark and under 670 nm light irradiation with various incident powers, indicating the considerable rectification characteristics in our homojunction device. The optical switch characteristic under zero bias (Figure 4b) further reveals the presence of the built-in field while the excitation power dependent photoresponse (Figure S8a) suggests the excellent linearity of the detector. In addition, the rise and fall time of the device, which are defined as the time taken for the response to increase from 10 percent to 90 percent of the peak and vice versa, are estimated to be 39 and 40 ms, respectively (Figure S8b). Figure 4c presents the spectral response under the bias of 0, 0.5, and -0.5 V, which reveals a high responsivity of 0.28 A W$^{-1}$ at zero bias under a 670 nm on-resonance excitation. In order to estimate the detectivity ($D^*$) of the polarimeter, we first measured the bandwidth of the device to be about 12 Hz at zero bias (Figure 4d). The noise power spectrum was then acquired through a Model SR770 FFT network analyzer (Figure 4e). At 12 Hz, the noise power density $S_n$ of our device is estimated to be $1.6 \times 10^{-29}$ A$^2$ Hz$^{-1}$. Subsequently, the detectivity was calculated by $D^* = R(A)^{1/2}/(S_n)^{1/2}$, where $R$ is the responsivity and $A$ is the device area.[42] Figure 4f exhibits the $D^*$ spectrum of the as-fabricated polarimeter with a peak value of $4.8\times10^{10}$ Jones at 670 nm, which is comparable to 2D material based polarization detectors and is thus beneficial to the high-performance full-Stokes polarimeter.[2,7,20,21]

In general, the metasurface structure is still the main method for the current on-chip full-Stokes polarimeters.[1,3,11,40] Metasurface based polarimeters usually exhibit lower measurement errors than our homojunction device, as shown in Table S1. Nevertheless, our polarimeter functions as the integration of a polarization recognition layer with a photodetector together, thus realizing the Stokes parameter measurement without relying on an extra filtering layer and the external bias, which would greatly simplify the device structure and reduce the device size and power consumption. Moreover, the homojunction device with atomically thin thickness can be easily integrated with other materials with higher integration density compared with the metasurface based Stokes polarimeter, thus presenting tremendous advantages in current integrated photonic chips.[32] For the origins of errors in our homojunction polarimeter, the limited CPGE response is the major factor, which greatly constrains the precision of parameter $S_3$. Although the current research on such filterless polarimeter is in the primary stage and there are still some problems to be solved, especially the complicated operation in the measurement process. Nevertheless, this drawback could be settled in future researches by improving the device structure, such as fabricating the homojunction arrays to match the device at different angles. Overall, our study provides a paradigm for the design of a filterless self-powered polarization detector, which can be applied to various materials and structures with the discrimination capability on both linearly polarized and circularly polarized light. Therefore, by reasonably optimizing the device structure and measurement system, we hope to further improve the performance and application range of the full-Stokes polarimeters.

**Conclusions**

In summary, we for the first time report on a self-powered filterless on-chip full-Stokes polarimeter based on a SL-MoS$_2$/FL-MoS$_2$ homojunction. Without relying on an extra filtering layer and the applied bias, the as-fabricated device can accurately estimate the SOP of incident polarized light through measuring the photocurrents at different rotation angles, and presents acceptable errors with the minimum values of 5%, 4.8% and 6.7% for parameters $S_1$, $S_2$ and $S_3$, respectively within the wavelength range from 650 to 690 nm. Our study provides a paradigm for the design of a filterless self-powered polarization detector with great simplicity, and this strategy can be applied to various materials and structures with the discrimination capability on both linearly polarized and circularly polarized light to construct filterless on-chip polarimeter. We believe our homojunction full-Stokes polarimeter would find promising potential applications in integrated optical and optoelectronic devices and chips benefiting from its small size and simple structure. Thus, our findings would open up a way towards 2D materials based polarization devices and motivate more investigations on the full-Stokes parameter detection and imaging.

## Methods

### Device Fabrication

For the SL-MoS$_2$/FL-MoS$_2$ homojunction device, the Cr/Au (5 nm/50 nm) electrodes on a 300 nm SiO$_2$/Si substrate were first fabricated by photolithography, thermal evaporation and lift-off process. Afterwards the substrates were treated by oxygen plasma for 180 s. Next, the FL-MoS$_2$ flake was exfoliated and transferred onto one of the Au stripes through a typical dry transfer method. Subsequently, the FL-MoS$_2$ flake was transferred onto the other strip of the electrodes and contacts with the previously transferred FL-MoS$_2$ flake to form the homojunction. Finally, the device was annealing in the argon protection environment at 270 °C for 1 h to clean the surface and improve the contact. For the full-Stokes polarimetric measurements, the homojunction device was aslant fixed on a rotation stage at a 45° angle to the horizontal plane. Two Au wires were used to contact the original Au electrodes to additional fixed electrodes for electrical measurement.

**Device Characterizations**

Both OM and fluorescence images were acquired on an Olympus BX53M fluorescence microscope. PL and Raman spectra were measured on a homebuilt Raman spectrometer system (Horiba iHR550) coupled with a liquid nitrogen cooled CCD. The PL measurements were conducted with a 600 g mm$^{-1}$ grating and excited by a 473 nm solid-state laser with a power of 10 W cm$^{-2}$. The Raman measurements were performed with an 1800 g mm$^{-1}$ grating and excited by a 532 nm solid-state laser with a power of 45 W cm$^{-2}$.

**Electrical and Optoelectronic Measurements**

A 250 W halogen tungsten lamp was used as the light source and was dispersed through a Horiba iHR320 monochromator. For the full-Stokes polarimetric measurements, the output monochromatic light passed through a polarizer, a $\lambda/2$ plate and a $\lambda/4$ plate to modulate the SOP and finally focused onto the homojunction device. For the photoconductivity measurements, the output light focused directly onto the device without passing through the polarization modulation module. The incident power density was measured by a Gentec pyroelectric detector. The acquired photocurrent was first amplified through a Stanford SR570 preamplifier and then measured with a Stanford SR830 lock-in amplifier. The response bandwidth were measured by adjusting the frequency of the chopper. The noise power spectrum was measured in a Model SR770 FFT network analyzer. The response time was acquired through a Tektronix MDO3032 digital oscilloscope. All measurements were performed in ambient conditions.

**Supplementary information**

The Supporting Information is available free of charge on the ACS Publications website.

**Authors' contributions**

C. F. and D. L. conceived the research and designed the experiments. C. F. carried out the experiments and wrote the manuscript. B. Z. characterized the samples. D. L. and J. L. revised the manuscript.

**Conflict of interest**

The authors declare no conflict of interest.

**Acknowledgements**

We acknowledge the support from National Key Research and Development Program of China (2018YFA0704403), NSFC (61674060) and the Innovation Fund of WNLO.

**References**

(1) Arbabi, E.; Kamali, S. M.; Arbabi, A.; Faraon, A. *ACS Photonics* **2018**, *5*, 3132-3140.

(2) Liu, F.; Zheng, S.; He, X.; Chaturvedi, A.; He, J.; Chow, W. L.; Mion, T. R.; Wang, X.; Zhou, J.; Fu, Q.; Fan, H. J.; Tay, B. K.; Song, L.; He, R.-H.; Kloc, C.; Ajayan, P. M.; Liu, Z. *Adv. Funct. Mater.* **2016**, *26*, 1169-1177.

(3) Lee, K.; Yun, H.; Mun, S.-E.; Lee, G.-Y.; Sung, J.; Lee, B. *Laser Photonics Rev.* **2018**, *12*, 1700297.

(4) Rubin, N. A.; D'Aversa, G.; Chevalier, P.; Shi, Z.; Chen, W. T.; Capasso, F. *Science* **2019**, *365*, eaax1839.

(5) Fang, C.; Xu, M.; Ma, J.; Wang, J.; Jin, L.; Xu, M.; Li, D. *Nano Lett.* **2020**, *20*, 2339-2347.

(6) Ji, C.; Dey, D.; Peng, Y.; Liu, X.; Li, L.; Luo, J. *Angew. Chem. Int. Ed. Engl.* **2020**, *59*, 18933-18937.

(7) Lai, J.; Liu, X.; Ma, J.; Wang, Q.; Zhang, K.; Ren, X.; Liu, Y.; Gu, Q.; Zhuo, X.; Lu, W.; Wu, Y.; Li, Y.; Feng, J.; Zhou, S.; Chen, J. H.; Sun, D. *Adv. Mater.* **2018**, *30*, 1707152.

(8) Li, D.; Zhang, J.; Zhang, Q.; Xiong, Q. *Nano Lett.* **2012**, *12*, 2993-2999.

(9) Basiri, A.; Chen, X.; Bai, J.; Amrollahi, P.; Carpenter, J.; Holman, Z.; Wang, C.; Yao, Y. *Light Sci. Appl.* **2019**, *8*, 78.

(10) Li, L.; Wang, J.; Kang, L.; Liu, W.; Yu, L.; Zheng, B.; Brongersma, M. L.; Werner, D. H.; Lan, S.; Shi, Y.; Xu, Y.; Wang, X. *ACS Nano* **2020**, *14*, 16634-16642.

(11) Jung, M.; Dutta-Gupta, S.; Dabidian, N.; Brener, I.; Shcherbakov, M.; Shvets, G. *ACS Photonics* **2018**, *5*, 4283-4288.

(12) Balthasar Mueller, J. P.; Leosson, K.; Capasso, F. *Optica* **2016**, *3*, 42-47.

(13) Tu, X.; McEldowney, S.; Zou, Y.; Smith, M.; Guido, C.; Brock, N.; Miller, S.; Jiang, L.; Pau, S. *Appl. Opt.* **2020**, *59*, G33-G40.

(14) Yan, C.; Li, X.; Pu, M.; Ma, X.; Zhang, F.; Gao, P.; Liu, K.; Luo, X. *Appl. Phys. Lett.* **2019**, *114*, 161904.

(15) Guan, X.; Yu, X.; Periyanagounder, D.; Benzigar, M. R.; Huang, J. K.; Lin, C. H.; Kim, J.; Singh, S.; Hu, L.; Liu, G.; Li, D.; He, J. H.; Yan, F.; Wang, Q. J.; Wu, T. *Adv. Opt. Mater.* **2020**, *101*, 2001708.

(16) Han, G. H.; Duong, D. L.; Keum, D. H.; Yun, S. J.; Lee, Y. H. *Chem. Rev.* **2018**, *118*, 6297-6336.

(17) Liu, C.; Chen, H.; Wang, S.; Liu, Q.; Jiang, Y. G.; Zhang, D. W.; Liu, M.; Zhou, P. *Nat. Nanotechnol.* **2020**, *15*, 545-557.

(18) Liu, Y.; Weiss, N. O.; Duan, X.; Cheng, H.-C.; Huang, Y.; Duan, X. *Nat. Rev. Mater.* **2016**, *1*, 16042.

(19) Wang, Y.; Ding, K.; Sun, B.; Lee, S.-T.; Jie, J. *Nano Res.* **2016**, *9*, 72-93.

(20) Guo, Q.; Pospischil, A.; Bhuiyan, M.; Jiang, H.; Tian, H.; Farmer, D.; Deng, B.; Li, C.; Han, S. J.; Wang, H.; Xia, Q.; Ma, T. P.; Mueller, T.; Xia, F. *Nano Lett.* **2016**, *16*, 4648-4655.

(21) Zhou, Z.; Long, M.; Pan, L.; Wang, X.; Zhong, M.; Blei, M.; Wang, J.; Fang, J.; Tongay, S.; Hu, W.; Li, J.; Wei, Z. *ACS Nano* **2018**, *12*, 12416-12423.

(22) Xia, J.; Yan, J.; Shen, Z. X. *FlatChem* **2017**, *4*, 1-19.

(23) Tan, C.; Zhang, H. *Chem. Soc. Rev.* **2015**, *44*, 2713-2731.

(24) Glazov, M. M.; Ivchenko, E. L.; Wang, G.; Amand, T.; Marie, X.; Urbaszek, B.; Liu, B. L. *Phys.*

*Status Solidi. (b)* **2015**, *252*, 2349-2362.

(25) Chen, M.; Lee, K.; Li, J.; Cheng, L.; Wang, Q.; Cai, K.; Chia, E. E. M.; Chang, H.; Yang, H. *ACS Nano* **2020**, *14*, 3539-3545.

(26) Eginligil, M.; Cao, B.; Wang, Z.; Shen, X.; Cong, C.; Shang, J.; Soci, C.; Yu, T. *Nat. Commun.* **2015**, *6*, 7636.

(27) Liu, L.; Lenferink, E. J.; Wei, G.; Stanev, T. K.; Speiser, N.; Stern, N. P. *ACS Appl. Mater. Inter.* **2019**, *11*, 3334-3341.

(28) Quereda, J.; Ghiasi, T. S.; You, J. S.; van den Brink, J.; van Wees, B. J.; van der Wal, C. H. *Nat. Commun.* **2018**, *9*, 3346.

(29) Cha, S.; Noh, M.; Kim, J.; Son, J.; Bae, H.; Lee, D.; Kim, H.; Lee, J.; Shin, H. S.; Sim, S.; Yang, S.; Lee, S.; Shim, W.; Lee, C. H.; Jo, M. H.; Kim, J. S.; Kim, D.; Choi, H. *Nat. Nanotechnol.* **2018**, *13*, 910-914.

(30) Luo, W.-M.; Shao, Z.-G.; Qin, X.-F.; Yang, M. *Physica E.* **2020**, *115*, 113714.

(31) Rasmita, A.; Jiang, C.; Ma, H.; Ji, Z.; Agarwal, R.; Gao, W.-b. *Optica* **2020**, *7*, 1204-1208.

(32) Duan, X.; Wang, C.; Pan, A.; Yu, R.; Duan, X. *Chem. Soc. Rev.* **2015**, *44*, 8859-8876.

(33) Sun, L.; Yan, J.; Zhan, D.; Liu, L.; Hu, H.; Li, H.; Tay, B. K.; Kuo, J. L.; Huang, C. C.; Hewak, D. W.; Lee, P. S.; Shen, Z. X. *Phys. Rev. Lett.* **2013**, *111*, 126801.

(34) Liu, X.; Chanana, A.; Huynh, U.; Xue, F.; Haney, P.; Blair, S.; Jiang, X.; Vardeny, Z. V. *Nat. Commun.* **2020**, *11*, 323.

(35) McIver, J. W.; Hsieh, D.; Steinberg, H.; Jarillo-Herrero, P.; Gedik, N. *Nat. Nanotechnol.* **2011**, *7*, 96-100.

(36) Yuan, H.; Wang, X.; Lian, B.; Zhang, H.; Fang, X.; Shen, B.; Xu, G.; Xu, Y.; Zhang, S. C.; Hwang, H. Y.; Cui, Y. *Nat. Nanotechnol.* **2014**, *9*, 851-857.

(37) Xu, S.-Y.; Ma, Q.; Shen, H.; Fatemi, V.; Wu, S.; Chang, T.-R.; Chang, G.; Valdivia, A. M. M.; Chan, C.-K.; Gibson, Q. D.; Zhou, J.; Liu, Z.; Watanabe, K.; Taniguchi, T.; Lin, H.; Cava, R. J.; Fu, L.; Gedik, N.; Jarillo-Herrero, P. *Nat. Phys.* **2018**, *14*, 900-906.

(38) Yu, J.; Xia, L.; Zhu, K.; Pan, Q.; Zeng, X.; Chen, Y.; Liu, Y.; Yin, C.; Cheng, S.; Lai, Y.; He, K.;

Xue, Q. *ACS Appl. Mater. Inter.* **2020**, *12*, 18091-18100.

(39) Guan, H.; Tang, N.; Xu, X.; Shang, L.; Huang, W.; Fu, L.; Fang, X.; Yu, J.; Zhang, C.; Zhang, X.; Dai, L.; Chen, Y.; Ge, W.; Shen, B. *Phys. Rev. B* **2017**, *96*, 241304.

(40) Afshinmanesh, F.; White, J. S.; Cai, W.; Brongersma, M. L. *Nanophotonics* **2012**, *1*, 125-129.

(41) Pors, A.; Nielsen, M. G.; Bozhevolnyi, S. I. *Optica* **2015**, *2*, 716-723.

(42) Kang, D. H.; Pae, S. R.; Shim, J.; Yoo, G.; Jeon, J.; Leem, J. W.; Yu, J. S.; Lee, S.; Shin, B.; Park, J. H. *Adv. Mater.* **2016**, *28*, 7799-7806.

**Figures**

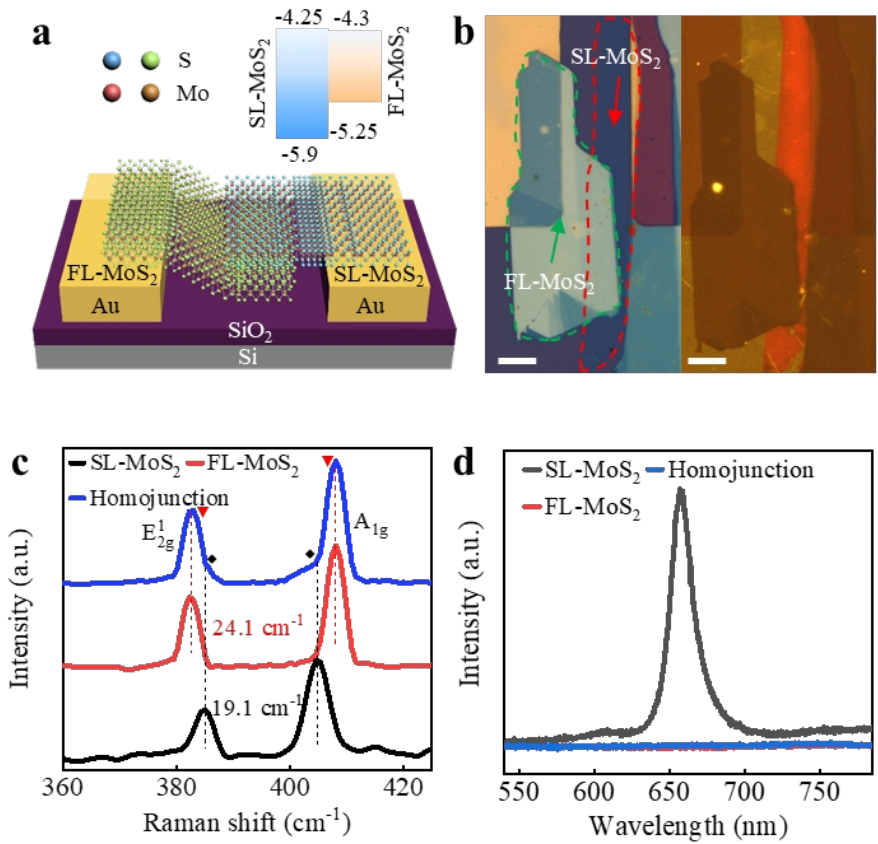

**Figure 1. The SL-MoS$_2$/FL-MoS$_2$ homojunction device. a** The schematic illustration of the homojunction device. The inset shows the energy band diagram of the homojunction. **b** OM (left) and fluorescent (right) images of the homojunction device. The scale bar is 10 μm. **c** Raman spectra in different regions of the homojunction device. Red triangles: phase belongs to FL-MoS$_2$, Black diamonds: phase belongs to SL-MoS$_2$. **d** PL spectra in different regions of the homojunction device.

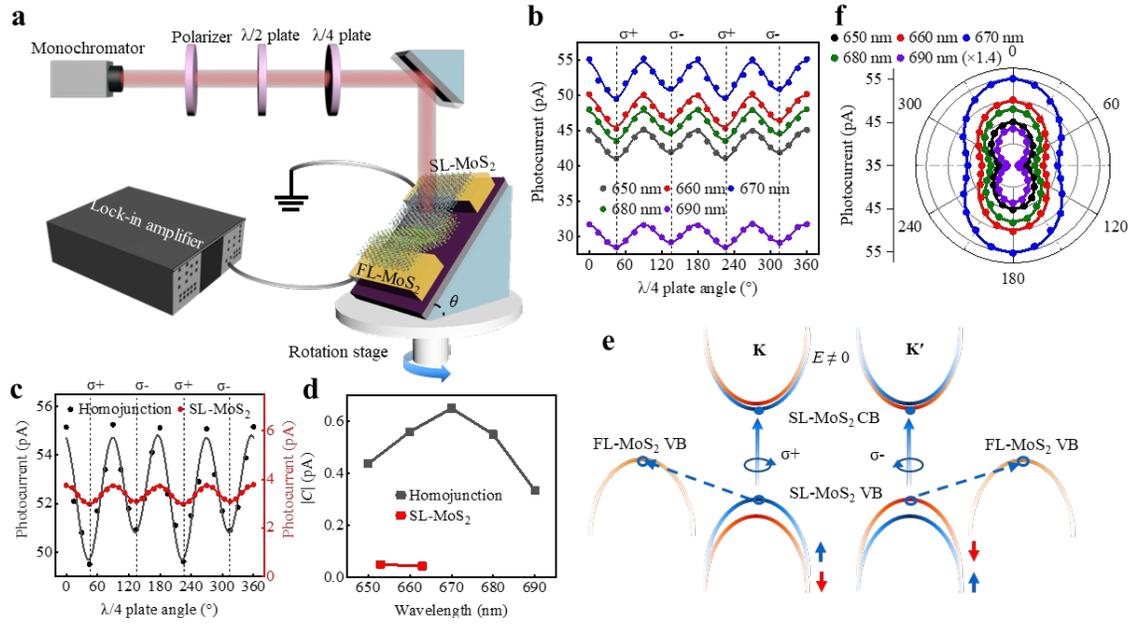

**Figure 2. The experimental setup and polarization responses of the device. a** The schematic illustration of the experimental setup and measurement configuration for the full-Stokes polarimetric measurements. The incidence angle $\theta$ is 45°. **b** The photocurrent as a function of the λ/4 plate angle at zero bias with various excitation wavelengths. The incident light power is 60 μW cm$^{-2}$. **c** Comparison of the helicity-dependent photocurrent between the homojunction device and the SL-MoS$_2$ at 0 V and -0.1 V, respectively. The excitation wavelength is 670 nm for the homojunction device and 653 nm for the SL-MoS$_2$, which corresponds to their photoresponse peak. The incident light power is 60 μW cm$^{-2}$. **d** Comparison of the helicity-dependent circular photocurrent parameter |C| between the homojunction device and the SL-MoS$_2$ with various excitation wavelengths. **e** The schematic illustration of the charge transfer effect during the helicity-dependent spin excitation process of this homojunction. The ultrafast transfer of holes enables the significant reduction of hole current, thus greatly enhanced the CPGE current. **f** Linear polarization-resolved photocurrent at zero bias with various excitation wavelengths.

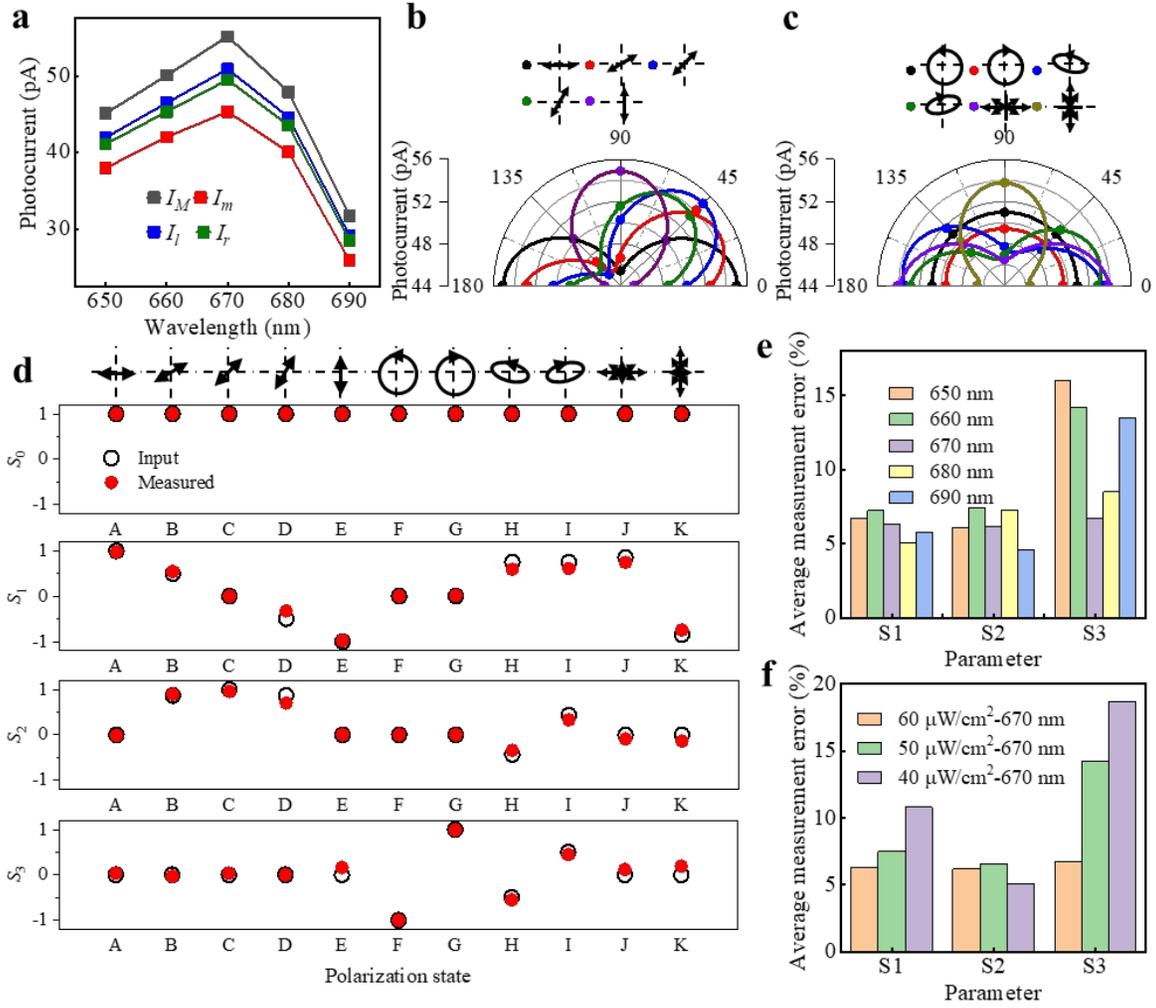

**Figure 3. The SOP measurement results of the device. a** The device parameters $I_M$, $I_m$, $I_l$ and $I_r$ at zero bias under various excitation wavelengths. The incident light power is 60 μW cm$^{-2}$. **b, c** Measured photocurrent under eleven different input polarized light excitation when the rotation stage is set to 0°, 45°, 90°, 135° and 180°. The incident light is 670 nm at a power of 60 μW cm$^{-2}$. **d** The input (black hollow circles) and measured (red solid circles) Stokes parameters ($S_0$-$S_3$) under different input polarized light excitation with the polarization ellipses illustrated at the top. **e, f** The average measurement errors of our full-Stokes polarimeter for **e** various excitation wavelengths and **f** various excitation powers.

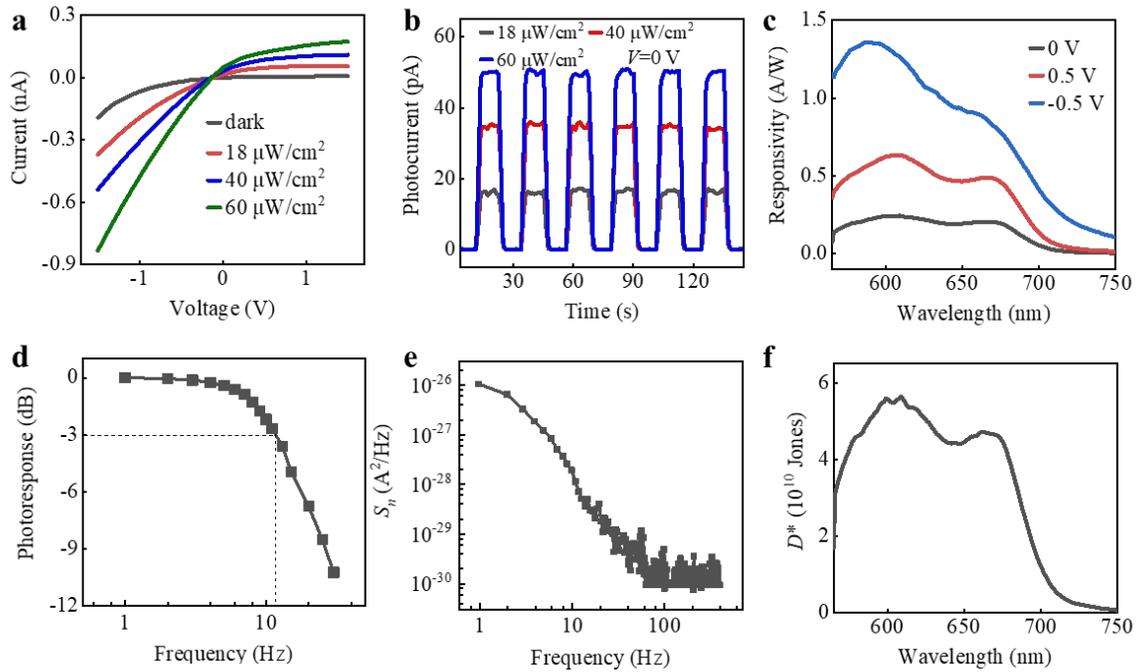

**Figure 4. The photodetection capability of the homojunction device. a** *I-V* curves of the homojunction device under 670 nm light irradiation under various incident powers. **b** Time-dependent photocurrent response of the homojunction device under switched-on/off 670 nm light irradiation with various incident powers at zero bias. **c** Responsivity spectra of the homojunction device under different biases. **d** Frequency response of the homojunction device at zero bias. **e, f** Noise power density ($S_n$) and detectivity ($D^*$) spectrum of the homojunction device at zero bias, respectively.

**TOC**

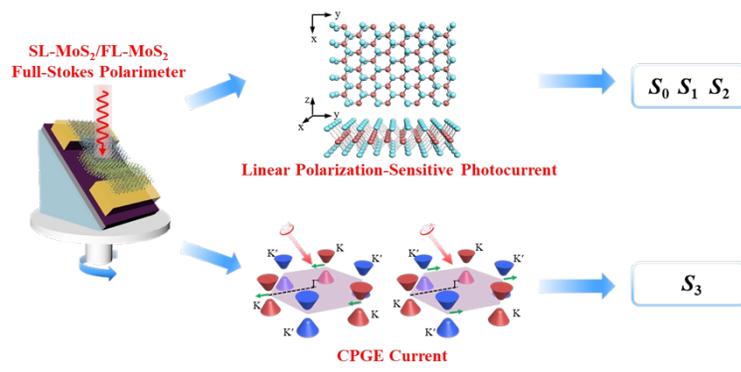